# Why Are Some Online Educational Programs Successful?: Student Cognition and Success

*Marissa Keech & Ashok Goel*
Design & Intelligence Laboratory, School of Interactive Computing
Georgia Institute of Technology, Atlanta, GA 30332
Marissa.gonzales@cc.gatech.edu , Ashok.Goel@cc.gatech.edu

**Abstract**

Massive Open Online Courses (MOOCs) once offered the promise of accessibility and affordability. However, MOOCs typically lack expert feedback and social interaction, and have low student engagement and retention. Thus, alternative programs for online education have emerged including an online graduate program in computer science at a major public university in USA. This program is considered a success with over 9000 students now enrolled in the program. We adopt the perspective of cognitive science to answer the question *why do only some online educational courses succeed?* We measure learner motivation and self-regulation in one course in the program, specifically a course on artificial intelligence (AI). Surveys of students indicate that students' self-reported assessments of self-efficacy, cognitive strategy use, and intrinsic value of the course are not only fairly high, but also generally increase over the course of learning. This data suggests that the online AI course might be a success because the students have high self-efficacy and the class fosters self-regulated learning.

## Introduction

MOOCs were once perceived as offering the potential for scalable education, making learning accessible, affordable and achievable to large segments of humanity. Although some MOOCs have been quite successful (Oakley, 2016), as MOOCs became widespread, the climate surrounding the promise of online education began to change. The MOOC approach to online learning, it is now said, often does not provide the structures and processes needed to support student success, for example, expert feedback, learning assistance, and social interaction, (Hollands & Tirthali, 2014; Kizilcec & Halawa, 2015). This results in online students bearing a high cognitive burden, making it difficult for many students to excel, with only a fraction of students actually completing their online classes (Anderson et al. 2014; Breslow et al. 2013).

Consequently, several academic institutions are now experimenting with alternate approaches to online education. For example, in 2014 a major public university in USA started an online program leading to a master's degree in computer science (Goel & Joyner 2016). Unlike traditional MOOCs, this is a highly-selective, low-cost, fully accredited program. Further, unlike traditional MOOCs, this model is widely perceived to be a significant educational success (Goodman et al. 2018). The number of students in the program has grown from just 200 since its inception in Spring 2014 to almost 9000 in Spring 2019, making it the largest graduate program in computer science in the US within five years This university has now started additional online educational programs explicitly modeled on the program in computer science. Other universities too have taken inspiration from the online program in computer science to develop their own programs.

The sharp contrast between the perceived success of the online program in computer science and the perceived non-success of many MOOCs courses raises a fundamental question: *why do only some online educational courses and programs succeed?* A number of hypotheses have been proposed to explain the discrepancy in the current context: for example, computer science is a highly specialized technical discipline, the student demographics in the program consists of highly specialized and educated individuals, the low but not insignificant financial investment made by the students acts as a push incentive for them to complete the program, the tangible financial benefits of completing the computer science program acts as a pull incentive, etc. All of these hypotheses seem plausible enough to require investigation.

In this paper, however, we adopt a socio-cognitive perspective and examine psychological aspects of student cognition in the online program in an attempt to answer the above question. The literature in educational psychology, and socio-cognitive theory relates perceived student self-efficacy and actual self-regulated learning with student success (Bandura, 1993; Pintrich et al. 1991). In particular, it proposes that (1) students with high perceived self-efficacy are more likely to be successful learners, and (2) learning environments that promote self-regulated learning in practice are more likely to result in student success. Of course, these potential explanations are not unrelated to earlier hypotheses: for example, it could be that highly educated students have higher self-efficacy than the general student population. Even so, the socio-cognitive explanations operationalize the earlier hypotheses and allow us to systematically study them. If student cognition is characterized by these motivational and cognitive constructs, then we should observe some evidence for them in the online program in computer science.

In this study, we examine the above hypotheses in the context of the online course in knowledge-based artificial intelligence (AI) (Goel & Joyner, 2016; Joyner, Goel & Isbell 2016). We estimate that the popular AI course has been taken by more than 4000 students in the computer science program; thus, about half of all students in the program have taken the AI course, making the course a good testbed for our cognitive explanations. In particular, we studied student demographics and student reported measures of motivation and self-regulation from two offering of this AI course, Spring 2017 and Fall 2017.

Below, we first determine whether or not students in the AI course express high or low motivation and self-regulation for learning. Second, we explore student demographics and what, if any, potential impact they have on our findings. Third, we explore how the student motivational and self-regulation measurements correlate to additional factors that contribute to determining student academic success; factors such as learning assessments and overall course performance. Finally, we examine our total findings and discuss if, perhaps, it is the students themselves, by way of exhibiting specific socio-cognitive constructs, that make the online AI course a success.

## Related Research

### Self-Efficacy and Self-Regulation in Learning

When it comes to motivational constructs, perceived self-efficacy is often considered one of, if not the, highest influence on academic achievement (Fernandez-Rio et al. 2017). The impact that self-efficacy has on student motivation makes it a powerful predictor of academic performance and effort (Schunk et al. 2002). The term self-efficacy "refers to beliefs in one's capabilities to organize and execute the courses of action required to produce given attainments," (Bandura, 1990), and includes not just the extent of those beliefs but the strength of them as well.

Multiple studies have found perceived self-efficacy to play a vital role in student academic success. Results from these studies indicate that perceived self-efficacy is influenced by many factors including learning environments (Wood & Bandura, 1989) the framing of subject content (Schunk, 1984), attitude (Collins et al. 1989), and academic anxiety (Meece et al. 1990). Students who have a low sense of perceived self-efficacy are made vulnerable to additional stressors and can become more erratic in analytic thinking [26]. Conversely, students who have a high sense of perceived self-efficacy are more resilient to negative forces and academic challenges (Bandura, 1993).

Additional research found that perceived self-efficacy plays a significant role in a student's ability to use appropriate cognitive strategies to enhance understanding, and to obtain help when necessary, thus impacting overall self-regulation (Bandura, 1993; Kizilcec et al. 2016; Zimmerman, 2000). This aligns well with Bandura's (Anderson, 2013) referring to self-efficacy as a self-regulation process.

Self-regulatory processes include those that help individuals take control of and reflect upon their learning behaviors. Processes for successful self-regulation also support the 'mastery goal orientation', an approach to learning that Zimmerman (Zimmerman et al. 1992; Zimmerman, 2000) states is characterized by learners who structure their learning experience around developing content knowledge and expertise.

Program environments and design that highlight self-regulatory processes and foster student motivation help reinforce positive self-beliefs that promote academic achievement (Kizilcec et al, 2016; Kilkarni et al. 2015). This is perhaps demonstrated best by examining how class design can impact whether or not students are able to successfully self-regulate. For instance, in traditional classroom-based instruction, the environment naturally lends itself to maintaining a fixed schedule and provides time for students to seek assistance in developing strategies for successful learning. In contrast, large and asynchronous environments complicate matters due to a common lack of enforced structure, something necessary for fostering self-regulation skills (Lajoie et al. 2006).

There have been many studies in recent years that examine what it takes to be successful in online education. The principle finding from these studies is that some behaviors and beliefs act as predictors to student success in an online learning setting (Artino Jr. et al. 2009; Azevedo, 20015; Dabbagh & Kitsantas, 2004; Kizilcec & Halawa, 2015). The behaviors identified as having an impact academic performance and student success in online education include the use of learning strategies for increased understanding (cognitive strategies, critical thinking, reflection), and strong motivational beliefs (intrinsic value and self-efficacy).

Based on the research above, we posit that measuring specific motivational and self-regulation components of students in the online AI course will help us determine if there is a relationship between the online AI course being success and the type of cognition exhibited by the students. Successful online programs might very well be the result of increasingly motivated and educated students. The investigation begins with first modeling student cognition using motivational and self-regulation constructs, and briefly reviewing objective measurements of student performance.

### Determining Online Course Success

Since the online CS graduate degree program's launch in 2014, over 4000 students have taken the online AI course that was used in this study. During this period, multiple studies have found that students in the online course have about the same course completion ratio as the equivalent course for residential students (Goel & Joyner 2016, 2017; Goel 2019). It is noteworthy that the AI course for both online and residential sections uses the same contents, materials, assessments, and graders. The multiple studies also found that for all assessments in the AI course, the online students performed as well as, if not better, than students in the equivalent course for the residential section (*references suppressed*). It is noteworthy again that these results are almost diametrically opposite from many MOOCs.

## Methodology

### Research Questions

In this work, we measured student motivational and self-regulated components in the online AI class. From the findings we begin to understand what psychological aspects of student cognition may have an impact on their student performance. Similarly, we determine if student self-reported measurements in self-regulation and motivation change through the term and compare student performance (as measured by student grades on the learning assessments throughout the semester) to student self-regulation and motivation measurements. Our research questions are:

RQ1: *Do students in the online AI course have high self-efficacy and do they use self-regulation in their learning?*

RQ2: *Do student measures for each construct change from the beginning of the term to the end of the term?*

RQ3: *Do student measurements in self-efficacy and self-regulation correlate to one another?*

From these research questions, we generated several hypotheses that we will attempt to validate in the remainder of the paper.

**Hypotheses**

H1: *Students in the online AI class have high perceived self-efficacy.*

H2: *Students in the AI class use self-regulation in their learning within the course.*

H3: *The self-efficacy of the students is at least partially informed by the demographics of the online student population, which are different from the demographics of the residential student population.*

H4: *Student cognition for those enrolled in the online AI class is characterized by high measurements of psychological constructs and may contribute to why the course is considered a success.*

**Student Engagement Survey**

The survey we used for our study borrowed (3) subscales from the Motivated Strategies for Learning questionnaire (MSLQ) (Pintrich et al. 1991), an instrument used for measuring student perceived components widely accepted as influencing overall student performance in education. Pintrich and De Groot originally developed the MSLQ for their motivation and self-regulated components study (Pintrich & DeGroot, 1990). From their resulting instrument, we borrowed subscales for measuring: self-efficacy, intrinsic value, and cognitive strategy use. All three subscales were relatively agnostic to class design, whereas additional subscales in the MSLQ were determined as not-applicable to the class in which the survey would be used. For instance, one of the subscales was specifically on test anxiety, and another detailed the use of readings and study materials. However, the online AI class used in this study neither assigns explicit reading and study materials, nor does it have timed or proctored tests.

The first two sub-scales of the survey, (self-efficacy, intrinsic value) are motivational components, and the third (cognitive strategy) is a self-regulation component. We added a fourth subscale, 'Confidence in Teaching Support', not so much to determine student motivation and self-regulation as much as to gather important information on student perceptions of teaching support in the course. This subscale provides meaningful information to the study, given that student perceptions of support can influence their overall performance (Picciano, 2002), and because adequate support is a prominent issue in scalable education. The resulting instrument is a 24-statement survey with (4) subscales. Each sub-scale contained 5-7 prompts that could be ranked on a 7-point Likert scale, with 1 = "Not at all true of me" and 7 = "Very true of me." The order of the 24 statements in the survey was randomized to avoid statement and subscale priming.

**Participants**

The participants for this study were primarily graduate students enrolled in the online AI course. The survey was confidential and student responses were anonymized for data analyses.

There are (3) groups of participants in this study. The first group is made up of students enrolled in the online program and who took the online AI class in the Spring 2017 semester. The second group is made up of students enrolled in the online program who took the online AI class in the Fall 2017 semester. The third group is made of students enrolled in the same AI class, offered at the same university that runs the online program, but in their residential equivalent. Thus, the final group took the AI class on campus. The three groups are: (1) Spring 2017 Online Students, (2) Fall 2017 Online Students, and (3) Fall 2017 Campus Students. The two online groups are entirely graduate students, while the campus group is a half graduate, and half undergraduate. Undergraduates who enroll in the class are required to have taken several preliminary courses prior to their enrollment in the course due to the advanced class content. Additional information can be found in earlier works (Goel & Joyner 2016; Joyner, Goel & Isbell 2016; Goel 2019).

All students were enrolled in the same AI class, meaning all groups used the class instructor, materials and assessments. This was possible because the AI class is offered through both the online and residential computer science programs: both programs use the same video lectures, educational tools and software, and utilize the same learning assessments and materials. We have included data from the residential program's campus-based version of the AI class to help put our findings from the online AI class into context, and to see if our findings were present in the campus-based AI class as well.

Demographic information from student participants in Spring 2017 Online and Fall 2017 Online groups is shown in Table 1, and demographic information for Fall 2017 Online and Fall 2017 Campus groups is shown in Table 2.

**Table 1. Demographic differences between online students in the 2017 Spring and Fall terms.**

|  | Online Spring 2017 | Online Fall 2017 |
|---|---|---|
| Age | <24: 19.0%<br>25-34: 58.8%<br>>35: 22.2% | <24: 17.9%<br>25-34: 59.0%<br>>35: 23.1% |
| Gender | Female: 13.6%<br>Male: 86.4% | Female: 10.3%<br>Male: 88.9% |
| Highest Level of Prior Education | Bachelor's: 80.6%<br>Master's: 15.1%<br>Doctoral: 4.3% | Bachelor's: 76.9%<br>Master's: 19.7%<br>Doctoral: 3.4% |
| Years of Programming Experience | <4: 23.7%<br>4-10: 53.4%<br>10-15: 13.3%<br>>15: 9.3% | <4: 22.2%<br>4-10: 53.0%<br>10-15: 14.5%<br>>15: 10.3% |

**Table 2. Demographic differences between online and residential students in the 2017 Fall terms.**

|  | Campus Fall 2017 | Online Fall 2017 |
|---|---|---|
| Age | <24: 73.0%<br>25-34: 23.1%<br>>35: 3.9% | <24: 17.9%<br>25-34: 59.0%<br>>35: 23.1% |
| Gender | Female: 23.07%<br>Male: 76.93% | Female: 10.3%<br>Male: 88.9% |
| Highest Level of Prior Education | Bachelor's: 88.5%<br>Master's: 11.5%<br>Doctoral: 0% | Bachelor's: 76.9%<br>Master's: 19.7%<br>Doctoral: 3.4% |
| Years of Programming Experience | <4: 88.5%<br>4-10: 11.5%<br>10-15: 0%<br>>15: 0% | <4: 22.2%<br>4-10: 53.0%<br>10-15: 14.5%<br>>15: 10.3% |

It is clear from the demographic data that while both online groups appear to be relatively comparable in terms of demographics, the same cannot be said for online students and campus students. The differences in population from the online and campus students is likely a product of the online environment, rather than a confound. A majority of the students the online groups identified as mature adults, and we know from previous research (Goel & Poleppeddi, 2016) that many of the students enrolled in the online program have external responsibilities, such as family and fulltime jobs, that impact the cycles they can dedicate to their education. As such, the online environment may hold more appeal than campus-based education to certain populations because it can better accommodate external forces and demands. This coincides with the general belief that online education would benefit groups of individuals who might otherwise have been unable to seek formal education.

## Experiments and Results

The survey was administered in a pre-post design, first at the beginning-of-term (BoT) and then again at the end-of-term (EoT) for each 16-week semester. We waited until the end of the second week in each semester to administer the first release of the survey to ensure us the student population in the class had stabilized. The survey was released again at the 15-week mark just before the final exam.

Response-rates are shown below in Table 3. For each semester, we have provided the response rate for the first release, second release and paired surveys. The 'Paired' section shows the number of students who took both the first and second surveys and it is these students whose data was subsequently used for this study. The ratio shown for each is the number of student responses / the total number of enrolled students in the semester.

**Table 3. First, Second and Paired survey response rates for each semester.**

|  | 1st Survey | 2nd Survey | Paired |
|---|---|---|---|
| Spring 2017 | 78/145 (53.8%) | 28/145 (19.3%) | 24/145 (17%) |
| Online Fall 2017 | 111/253 (43.9%) | 112/253 (44.3%) | 73/253 (29%) |
| Campus Fall 2017 | 31/83 (37.3%) | 24/83 (28.9%) | 17/83 (20%) |

### Analysis of Independent Variables

We used only paired surveys from each group for this analysis. Any significant outliers were considered on a case-by-case basis before being removed from each sample. Examples of the kinds of records removed from the sample include students who did not complete all questions in the survey, or had corresponding grade data that could not be used (such as a student who was missing the mid-term and final exams in the course), therefore resulting in inadequate data to make any comparisons between their survey measurements and their grade data. In Spring 2017 Online, there were no significant outliers; in Fall 2017 Online, there were 3 significant outliers; in Fall 2017 Campus, there were 2 significant outliers. The final number of participants used in the study were: Spring 2017 Online (N=24), Fall 2017 Online (N=70) and Fall 2017 Campus (N=15). Response rates based on these numbers and the total number of enrolled students for each group in each semester are 17%, 28% and 18% respectively.

Unique student responses were averaged for each of the construct subscales measured by the survey. For example, the self-efficacy component contained five statements, and for each student these five statements were averages into a single self-efficacy score. This resulted in four unique construct scores per student in each group.

A paired-samples t-test was used to determine whether there was a statistically significant mean difference from BoT and EoT measurements for student self-efficacy, cognitive strategy use, intrinsic value, and confidence in teaching support. Data are the mean ± standard deviation unless otherwise stated. For each construct, we performed a boxplot test and Shapiro-Wilk's test to check for normal distribution. All analysis was run using SPSS. Additionally, we ran a Pearson's correlation test to identify correlations between the self-efficacy construct and the remaining three survey constructs.

**Student Engagement Survey Results**

All data results are shown in Tables 4, 5, and 6 for the Spring 2017 Online, Fall 2017 Online and Fall 2017 Campus groups respectively. Likewise, results from the Pearson's Correlation tests are shown in Tables 7, 8 and 9 for Spring 2017 Online, Fall 2017 Online and Fall 2017 Campus respectively.

In the following tables, SE is self-efficacy, CS is cognitive strategy, IV is intrinsic value, and CiTS is confidence in teaching support. Mean and standard deviation are shown end-of-Term listed first, Beginning-of-Term second. Columns for mean difference, standard error, p-value, t-statistic and effect size are calculated on the difference between pre-and-post.

**Table 4. Findings from Spring 2017 Online Student Engagement Survey.**

| Spring 2017 Online Student-Engagement Survey | | | | | | |
|---|---|---|---|---|---|---|
| | Mean ± Std. | Mean Diff. | Std. Err. | p-Val | t-Stat | Eff. Size |
| SE | 5.88± .67 5.55± .83 | .325 | .147 | .038 | 2.206 | .450 |
| CS | 5.72± .81 5.58± .92 | .143 | .167 | .398 | .861 | .176 |
| IV | 5.86±1.05 5.74± .88 | .127 | .160 | .435 | .794 | .162 |
| CiTS | 6.31± .57 6.14± .83 | .167 | .139 | .242 | 1.202 | .245 |

**Table 5. Findings from Fall 2017 Online Student Engagement Survey.**

| Fall 2017 Online Student-Engagement Survey | | | | | | |
|---|---|---|---|---|---|---|
| | Mean ± Std. | Mean Diff. | Std. Err. | p-Val | t-Stat | Eff. Size |
| SE | 5.77± .76 5.46± .81 | .280 | .099 | .003 | 3.08 | .368 |
| CS | 5.65± .78 5.62± .78 | .031 | .085 | .721 | .359 | .043 |
| IV | 5.75± .97 6.01± .77 | -.260 | .098 | .010 | -2.632 | -.315 |
| CiTS | 6.10± .80 5.85± .83 | .250 | .087 | .006 | 2.838 | .340 |

**Table 6. Findings from Fall 2017 Campus Student Engagement Survey.**

| Fall 2017 Campus Student-Engagement Survey | | | | | | |
|---|---|---|---|---|---|---|
| | Mean ± Std. | Mean Diff. | Std. Err. | p-Val | t-Stat | Eff. Size |
| SE | 5.64± .97 5.91± .73 | -.267 | .256 | .005 | 2.888 | .350 |
| CS | 5.54±1.03 5.73± .57 | -.195 | .188 | .357 | .928 | .115 |
| IV | 5.84±1.23 6.07± .89 | -.229 | .193 | .025 | -2.294 | .28 |
| CiTS | 5.93±1.04 6.39± .46 | -.463 | .213 | .005 | 2.879 | .35 |

**Table 7. Pearson's Correlation between Self-Efficacy and the remaining constructs. EoT results for Spring 2017 online.**

| Spring 2017 Online Self-Efficacy Correlation | | | |
|---|---|---|---|
| | Pearson's Correlation | Coefficient Value | Strength of Association |
| CS | .522 | .009 | Strong + |
| IV | .439 | .032 | Moderate + |
| CiTS | .642 | .001 | Strong + |

**Table 8. Pearson's Correlation between Self-Efficacy and the remaining constructs. EoT results for Fall 2017 online.**

| Fall 2017 Online Self-Efficacy Correlation | | | |
|---|---|---|---|
| | Pearson's Correlation | Coefficient Value | Strength of Association |
| CS | .403 | .001 | Moderate + |
| IV | .491 | <.0001 | Moderate + |
| CiTS | .531 | <.0001 | Strong + |

**Table 9. Pearson's Correlation between Self-Efficacy and the remaining constructs. EoT results for Fall 2017 campus.**

| Fall 2017 Campus Self-Efficacy Correlations | | | |
|---|---|---|---|
| | Pearson's Correlation | Coefficient Value | Strength of Association |
| CS | .585 | .022 | Strong + |
| IV | .518 | .048 | Strong + |
| CiTS | .660 | .007 | Strong + |

## Discussion

### Research Question Discussion

First, we look at our research questions individually, and then reflect on our findings overall.

**RQ1:** Do students in the online AI course have high self-efficacy and do they use high-levels self-regulation in their learning?

This survey, which measured student motivational and self-regulatory components, showed that students in all groups (Spring and Fall 2017 Online, and Fall 2017 Campus) had relatively high measures in all survey constructs at both the BoT and the EoT. We consider the average across all constructs ($M$=5.71, $M$=5.74, $M$=6.03 for Spring and Fall online, and Fall Campus respectively) for each group at the BoT as 'high' measurements based on Pintrich and DeGroot's finding a mean of 5.41 in their original study on motivation and self-regulation in the classroom as high (Pintrich & DeGroot, 1990).

That participants initially reported high measurements in all constructs in each group, might suggest that all groups consisted of relatively confident and experienced students. This is supported by the demographic data. Given that the online program is a graduate-level program, we anticipated this. However, the Fall 2017 Campus group consists of both undergraduate and graduate students, and among the undergraduate participant students alone (N=7), we see a BoT overall construct mean of ($M$= 6.12).

**RQ2:** Do student measures in each construct in the AI course change from the beginning of the term to the end of the term?

In all groups, self-efficacy showed a statistically significant increase. Cognitive strategy use, while increasing in both terms from BoT to EoT, did not increase a statistically significant amount. Intrinsic value did increase in the Spring 2017 Online group, but not significantly. In both Fall 2017 groups, Online and Campus, the mean difference from BoT to EoT showed a statistically significant decrease in intrinsic value. Finally, while confidence in teaching support did not show a significant increase from BoT to EoT in the Spring 2017 Online group, it did show a statistically significant increase in both the Online and Campus Fall 2017 groups.

The statistically significant increase in self-efficacy across all groups could be the result of students getting the support necessary to influence and increase their self-confidence. Support such as formative feedback and guidance, methods present in both the online and campus versions of the AI course, could be partially responsible for this increase. Student self-efficacy can be fostered with positive engagement with learning materials and cognitive engagement. Positive course experiences, such as timely feedback, thriving discussion among peers, and constructive teacher-student relationships can help positively polarize the overall learning experience, and by extension student self-beliefs.

Despite the statistically significant decrease in intrinsic value in both Fall 2017 groups, the measurements were still relatively high. Possible explanations for this change could be that students' preconceived notions of the AI class were violated by the end of the semester. Students may have had specific expectations of the class or the class objectives, and if those expectations were not met, it might account for the changes found in the EoT results. This measurement does not necessarily mean that students do not value their class experience; rather, it suggests that by the end of the course, students may have discovered a different kind of knowledge than what they were originally expecting and therefore would have to adjust their earlier expected applications for such knowledge.

In the Spring 2017 online group, our data did not show a statistically significant increase in student confidence in teaching support. However, there did appear to be a statistically significant increase for both Fall 2017 groups. There are many factors that influence a student's confidence in their teaching staff and support. Among them, the level of student-teacher interactions either privately or publicly through the online discussion forum, or perhaps student confidence in teaching support raised because they felt the teaching staff was quick to respond to questions and concerns, and quickly returned graded assignments. Additional research is in-progress as we attempt to gain further insight into the specific elements of the online course that influenced these findings.

**RQ3:** Do student measurements in self-efficacy and self-regulation correlate to one another?

Results from our correlation analysis showed that across all groups, use of cognitive strategies, intrinsic value, and confidence in teaching support all significantly and positively correlated with self-efficacy.

These findings correlate to those of Pintrich and DeGroot's study (Pintrich & DeGroot, 1990) where they found self-efficacy to be positively related to components of cognitive engagement. One way of interpreting the relationship between self-efficacy and use of cognitive strategies is that students who reported perceptions of high self-capability were more likely to report use of cognitive strategies, which as Pintrich and DeGroot found, also implied that students were self-regulating.

There does appear to be a correlation between intrinsic value and self-efficacy in all groups, and this too corresponds with what Pintrich and DeGroot found in their study. Similarly, a relationship between intrinsic value and use of cognitive strategies was found in both Fall 2017 groups. Considering these two points, it suggests that even though the class wasn't what they were expecting, it didn't negatively impact their confidence that they could do well, and that students who were motivated to learn the material for more than extrinsic reasons were more likely to be cognitively engaged in learning.

Finally, a relationship between self-efficacy and confidence in teaching assistance suggests that students who were confident in the support they received in the class were more likely to be confident in their ability to perform well on class materials and assignments.

## Hypotheses Discussion

**H1** *Students in the online AI class have high perceived self-efficacy.*

As the data indicates, students in the online AI class do have higher measurements of both the motivational and self-regulation components. Not all constructs increased significantly from the BoT to the EoT; however, all constructs were rated relatively high in both terms throughout the semester. This suggests that the students' cognition is characterized by high self-confidence as well as and high self-regulation. It also suggests that the students are self-motivated and driven to use strategies that positively impact their academic performance, such as goal setting, critical thinking, meta-cognition and reflection.

**H2** *Students in the AI class use self-regulation in their learning within the course.*

The correlation we found between self-efficacy and the other constructs (intrinsic value, cognitive strategy use, and confidence in teaching support) supports the findings of the literature that a relationship exists between motivation for learning and self-regulated learning. Students whose cognition is characterized by high self-efficacy and who regularly use cognitive strategies in their studies are more likely to be successful in their online academic endeavors.

Our findings suggest that the students in the online AI course are self-regulating and have high self-efficacy, which may be indicative of positive cognitive engagement with the course. As the literature suggests, and our results from additional research pending publication supports, these findings may account for increased academic performance of the students.

**H3** *The self-efficacy of the students is at least partially informed by the demographics of the online student population, which are different from the demographics of the residential student population.*

Whereas above we briefly discussed that increased measurements in the various psychological constructs could suggest that students are cognitively engaged with the course, here we will briefly discuss an alternative explanation. The findings of our research may suggest that students exhibit these high measurements in motivation and self-regulation because of the group's characteristics. This hypothesis is supported by the findings of the demographic data for both online groups, which shows that a majority of students already obtained at least one college degree or had between four and ten years of programming experience. Given that the online program is a graduate-level program, we anticipated this. However, it is worth noting that simply being a graduate student does not translate to automatically being confident in one's work or having positive cognitive engagement with course.

The survey results suggest that students regularly used self-regulation practices and had positive self-beliefs such as self-efficacy and intrinsic value. A majority of the students in the two terms analyzed in this study self-identified as mature adults, and we know from previous research (Joyner, Goel & Isbell 2016) that many of the students enrolling in the online program have external obligations, such as family and fulltime jobs, that impact the amount of time and attention they can dedicate to their education.

These factors influence a student's perceived learning experience and cognition, impacting their academic performance. Our findings suggest that student cognition in the online AI course is influenced by their demographic characteristics: older, more experienced, and motivated by external factors and forces.

**H4** *Student cognition for those enrolled in the online AI class is characterized by high measurements of psychological constructs and may contribute to why the course is considered a success.*

Again, we expect that the student's high self-efficacy scores in the online groups are related in part to the demographics of the students. As Tables 1 and 2 indicate, the online students tend to be older, have higher levels of prior education, and more programming experience than their residential students in the course. Many students in the online program are successful professionals with many years of experience in the information technology industry. Hence, the high self-efficacy scores.

More interesting is the increase in the perceived self-efficacy of the students over the term of the online AI class, and the relationship between self-efficacy measures and the other constructs (see Tables 7-9). Findings from additional research pending publication show that students in the Spring 2017 Online group showed a decrease in averages on their projects over the duration of the term. Normally this might call for concern, but when we consider that the projects in the class increase in complexity from one to the next, it is expected that students might experience a shift in performance due to increasingly complex and challenging tasks. In this same group we saw an increase in self-efficacy, and a strong positive correlation with confidence in teaching support and use of cognitive strategies. This implies that students might feel confident in their abilities to perform well on the class tasks because they are cognitively engaged in the course materials, because they confident in their ability to receive support from the teaching staff if they should need it, or because they carry the necessary characteristics for excelling in the coarse: (1) experience, (2) motivation both intrinsic and extrinsic, and (3) developed strategies for success.

Given that the students themselves carry these indicators for success, it may very well be that the course is successful not solely because of the curricula or design, but because the students themselves are successful. The students who enroll in the online course are characterized by cognitive engagement and unique relationships between motivation and self-regulation, as well as external forces, that drive them to perform well in the course.

## Conclusion

These findings provide preliminary evidence that the online AI course is represented by students who have high measures of motivation and self-regulation for learning. While it is not possible to generalize from one course to the online graduate program in computer science as a whole, the above results indicate that at least one popular course in the online program has many successful students, which themselves may be the reason for the courses success. These results invite additional research in other courses in the online program, as well as into additional factors for considering student cognition for success.

## Acknowledgements

We would like to thank Georgia Institute of Technology for their assistance on providing the demographic data used in this paper, as well as the entire Design & Intelligence lab for their support and helpful comments.